\documentclass[%
 pre,
aps,
 amsmath,amssymb,
 reprint,%
]{revtex4-1}

\usepackage{graphicx}
\usepackage{dcolumn}
\usepackage{bm}

\usepackage[utf8]{inputenc}
\usepackage[T1]{fontenc}
\usepackage{mathptmx}

\newcommand{\pd}{\partial}
\def\d{\mathrm{d}}
\renewcommand{\b}{\boldsymbol}
\def\cO{\mathcal{O}}
\def\b#1{\mathbf{#1}}
\newcommand{\xibar}{\bar{\xi}}
\newcommand{\etabar}{\bar{\eta}}
\begin{document}
\title{Definition and properties of logopoles of all degrees and orders}
\author{Matt Majic} \email{mattmajic@gmail.com}
\author{Eric C. Le Ru} \email{eric.leru@vuw.ac.nz}

\affiliation{The MacDiarmid Institute for Advanced Materials and Nanotechnology,
School of Chemical and Physical Sciences, Victoria University of Wellington,
PO Box 600, Wellington 6140, New Zealand}

\date{\today}

\begin{abstract}
Logopoles are a recently proposed class of solutions to Laplace's equation with intriguing links to both solid spheroidal and solid spherical harmonics. They share the same finite line singularity with the former and provide a generalization of the latter as multipoles of negative order. In [Phys. Rev. Res. {\bf 1}, 033213 (2019)], we introduced and discussed the properties and applications of these new functions in the special case of axi-symmetric problems (with azimuthal index $m=0$). This allowed us to focus on the physical properties without the added mathematical complications. 
Here we expand these concepts to the general case $m\neq 0$. The chosen definitions are motivated to conserve some of the most interesting properties of the $m=0$ case. 
This requires the inclusion of Legendre functions of the second kind with degree $-m\leq n<m$ (in addition to the usual $n\geq |m|$) and we show that these are also related to the exterior spheroidal harmonics. 
We show that Logopoles can also be defined for $n\le m$, and discuss in particular logopoles of degree $n=-m$, which correspond to the potential of line segments of uniform polarization density.
\end{abstract}
\maketitle

\section{Introduction}

The most commonly used solutions to Laplace's equation are the so-called multipoles or solid spherical harmonics (SSHs) of the form $r^{-n-1}P_n^m(\cos\theta) e^{im\phi}$. Another widely applicable class of solutions are the prolate solid spheroidal harmonics (PSSHs) of the form $Q_n^m(\xi) P_n^m(\eta) e^{im\phi}$, where $P_n^m$ and $Q_n^m$ are the Legendre functions of the first and second kinds. One fundamental difference between these solutions is their singularity: a finite line for PSSHs as opposed to a single point for SSHs.
Both SSHs and PSSHs are usually only considered for integer order $m$ and integer degree with $n\geq|m|$ in order to have solutions with a bounded singularity (often in physics $n$ is termed the multipole \textit{order}, but we will reserve this term for $m$). Other solutions, such as the SSHs of the 2$^{\rm nd}$ kind (replacing $P_n^m$ by $Q_n^m$) are typically excluded because of their unbounded singularities. 

Another class of solutions dubbed Logopoles and denoted $L_n^m$ were introduced in Ref.~\cite{majic2019logopoles} and have similarities with both SSHs and PSSHs. They exhibit many properties similar to SSHs (recurrence relations, ladder operators), but  have a finite line singularity like PSSHs. In Ref.~\cite{majic2019logopoles}, only logopoles with $m=0$ (which are then axisymmetric) were considered. Promising applications of these new functions were discussed, but to fully realize their potential, it is necessary to extend this concept to $m\neq 0$. As summarized in Sec.~\ref{SecBackground}, logopoles can be introduced/defined from different perspectives: integral representation, series definitions in terms of SSHs and PSSHs, ladder operators. There are different possibilities to generalize these definitions to $m\neq 0$, and the one we propose conserves the most ideal properties of logopoles, especially the finite line singularity and the close link to SSHs. These definitions are presented in Sec.~\ref{secLogopoles1} and their resulting properties are derived and discussed in Sec.~\ref{secLogopoles2}.

In Ref.~\cite{majic2019logopoles}, extensive use was made of the SSHs of the 2$^{\rm nd}$ kind, $r^nQ_n(\cos\theta)$, by taking linear combinations which canceled out the singularities on the $z$-axis towards $z\rightarrow\pm\infty$, to create spheroidal harmonics and logopoles, which have bounded singularities. As discussed in Secs.~\ref{secLogopoles1} and \ref{secNewRelationship}, the same can be done for $m>0$, but we found that this extension requires harmonics containing $Q_n^m(\cos\theta)$ with $n<m$ and even $n<0$. Unlike $Q_n^m$ for $n\geq m$ which contain logarithmic terms, these are purely rational functions. This also led us to consider logopoles of negative degree $L_{-n}^m$, which are discussed in Sec.~\ref{negative}. These also have potential physical applications, for example logopoles with $n=-m$ correspond to uniform polarization densities on the focal segment. 
In addition, a new integral expression for the SSHs of the 2$^\text{nd}$ kind was presented in Ref.~\cite{majic2019logopoles} for $m=0$ and positive powers of $r$. This is extended to $m\geq0$ and negative powers of $r$ in Sec.~\ref{secLineInts}.
Finally we discuss in Sec.~\ref{SecStable} stable numerical schemes to compute logopoles of any degree and order.


\section{Background}
\label{SecBackground}

We first define the main notations and summarize the important definitions/properties of logopoles as introduced in Ref.~\cite{majic2019logopoles}.

\subsection{Coordinate systems}
\begin{figure}
\includegraphics[scale=.9]{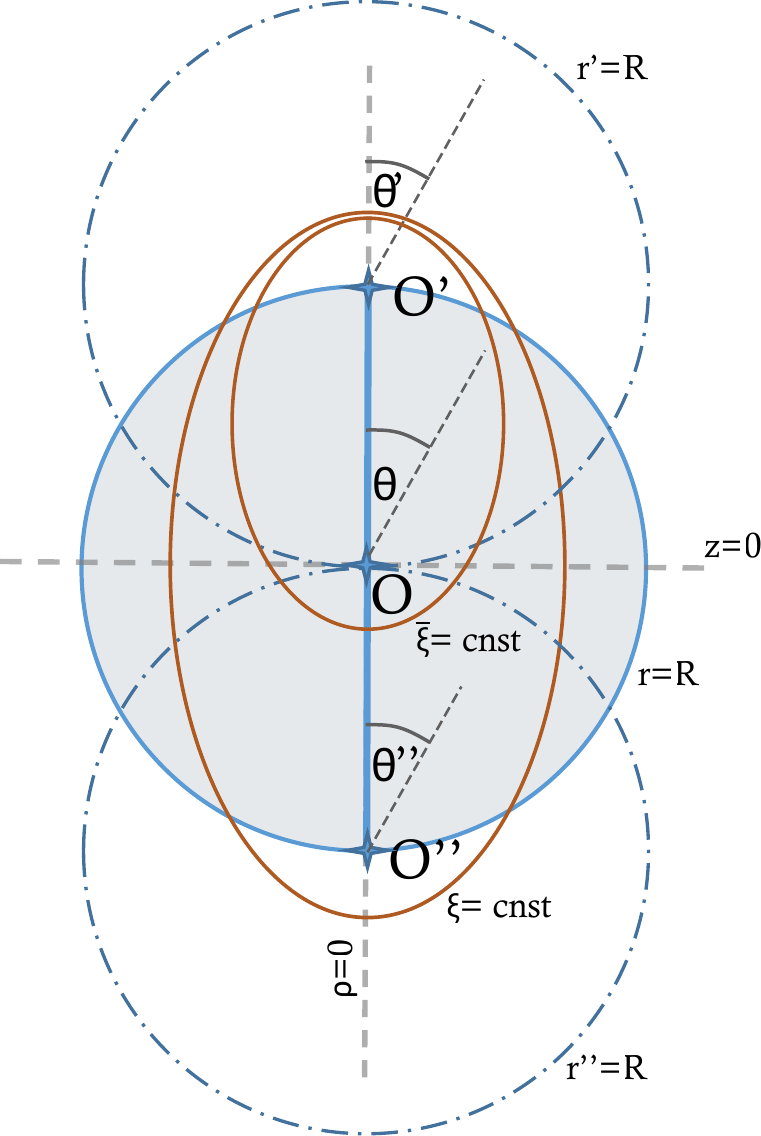}
\caption{Schematic of the offset spherical and spheroidal coordinate systems} \label{CoordinateSystems}
\end{figure}

We will use spherical coordinates $(r,\theta,\phi)$ (with $u\equiv\cos\theta$) centred at the origin O, cylindrical coordinates $(z,\rho,\phi)$, and define two translated coordinate frames, centred at O' and O'', offset by $+R$ and $-R$ along the $z$-axis (see figure \ref{CoordinateSystems}). These coordinates can be expressed as
\begin{align*}
\begin{array}{lll}
\rho'&=\rho=\sqrt{x^2+y^2}\\
z'&=z-R\\
r'&=\sqrt{\rho^2+(z-R)^2}\\
u'&=\cos\theta' = z'/r'
\end{array}
\qquad
\begin{array}{lll}
\rho''&=\rho\\
z''&=z+R\\
r''&=\sqrt{\rho^2+(z+R)^2}\\
u''&=\cos\theta'' = z''/r''
\end{array}
\end{align*}
We also define adimensional ``hat'' coordinates that are scaled by $R$, for example $\hat{r}'=r'/R$.\\
The conventional prolate spheroidal coordinates ($\xi$,$\eta$,$\phi$) are defined with focal line O''O': 
\begin{align}
\xi= \frac{r''+r'}{2R}, \qquad \eta=\frac{r''-r'}{2R}, \label{spheroidal coords}
\end{align}
and we will also use  offset spheroidal coordinates with focal line OO':
\begin{align}
\xibar=\frac{r+r'}{R},\quad \etabar=\frac{r-r'}{R}. \label{offset coords}
\end{align}

\subsection{Logopoles for $m=0$}

For convenience, we here reproduce the main formulae from Ref.~\cite{majic2019logopoles}.
First we define the external SSHs of the first kind $S_n^m$ and internal SSHs of the 2$^{\rm nd}$ kind $\tilde{S}_n^m$:
\begin{align}
S_n^m(\hat r,\theta,\phi)&=\hat r^{-n-1}P_n^m(\cos\theta)e^{im\phi}, \\
\tilde{S}_n^m(\hat r,\theta,\phi)&=\hat r^n Q_n^m(\cos\theta)e^{im\phi}.
\end{align}

Logopoles for $m=0$ can be defined either via an integral over a line of charge:
\begin{align}
L_n^0=&R\int_0^1\frac{ v^n\d v }{\sqrt{\rho^2+(z-Rv)^2}},\label{intL0}
\end{align}
or as an infinite series of multipoles:
\begin{align}
L_n^0(\hat{r},\theta,\phi)&=\sum_{k=0}^{\infty} \frac{S_k^0}{n+k+1}\label{LvsS0},
\end{align}
or as a finite sum of SSHs of the 2$^{\rm nd}$ kind:
\begin{align}
L_n^0= \tilde{S}^0_n - \sum_{k=0}^{n} \binom{n}{k}\tilde{S}_k^{0\prime}.
\label{LvsSQ0}
\end{align}
The prime means that the function is of coordinates centred at O': $\tilde S_k^{0\prime}=\tilde S^0_k(\hat r',\theta',\phi)$. 
From these equivalent definitions, one can show that $L_n^0$ can also be expressed as a finite sum
of PSSHs:
\begin{align}
L_n^0 = \sum_{k=0}^n \frac{2n!^2(2k+1)}{(n-k)!(n+k+1)!} Q^0_k(\xibar)P^0_k(\etabar)\label{LvsQPbar0},
\end{align}
and vice-versa:
\begin{align}
Q^0_n(\xibar)&P^0_n(\etabar) = \sum_{p=0}^n\frac{(-)^{p+n}(n+p)!}{2~p!^2(n-p)!} L^0_p. \label{QPbarvsL0}
\end{align}

Moreover,  the ladder operator for multipole degree $n$, $R\partial_z$, has the following effects:
\begin{align}
R\partial_z S^0_n&=-(n+1)S^0_{n+1}, \label{pdzS0}\\
R\partial_z \tilde{S}^0_n&=n\tilde{S}^0_{n-1}, \label{pdzSQ0}\\
R\pd_z L^0_n&=nL^0_{n-1}-S_0^{0\prime}. \label{L dif0}
\end{align}
The latter two equalities highlight the similarity between $\tilde{S}^0_n$ and logopoles.

Finally, combining these properties, we proved in Ref.~\cite{majic2019logopoles} a new relationship between PSSHs and SSHs of the 2$^{\rm nd}$ kind:
\begin{align}
Q^0_n(\xi)&P^0_n(\eta)=\sum_{k=0}^n\frac{(n+k)!}{2k!^2(n-k)!(2R)^k}\nonumber\\
&\times\left[(-)^{n+k}r''^kQ^0_k(\cos\theta'')-r'^kQ^0_k(\cos\theta')\right]. \label{newrel0} 
\end{align}

\section{Generalization of logopoles to $m\geq0$: Definitions}
\label{secLogopoles1}

One could conceive of several possible generalizations of a class of functions such as logopoles to include the index $m$, but the definition we will present stands out as being the most natural in many respects. 

\subsection{Definition via ladder operations}

Consider the angular momentum ladder operator
\begin{align}
\partial_+\equiv \pd_x+i\pd_y ~= \frac{\pd}{\pd x}+i\frac{\pd}{\pd y},
\end{align}
or in cylindrical and spherical coordinates:
\begin{align}
\partial_+=&e^{i\phi}\left[\partial_\rho+\frac{i}{\rho}\partial_\phi\right]\\
=&e^{i\phi}\left[\sin\theta\partial_r + \frac{\cos\theta}{r}\partial_{\theta} +\frac{i}{r\sin\theta}\partial_\phi\right].
\end{align}

Acting on spherical harmonics, $R\partial_+$ is a ladder operator for order $m$, explicitly \cite{1998NiceSummary}:
\begin{align}
R\partial_+S_n^m&=-S_{n+1}^{m+1}, \label{pd+S}\\
R\partial_+\tilde{S}_n^m&=-\tilde{S}_{n-1}^{m+1}. \label{pd+SQ}
\end{align}

Our proposed definition of logopoles for $m>0$ is via this operation:
\begin{align}
R\partial_+L_n^m&=-L_{n-1}^{m+1}.  \label{pd+L}
\end{align}
We will see how this definition applies naturally to many other formulae for logopoles.

\subsection{Integral representation}
 \label{line ints}

Logopoles for $m=0$ are easily interpreted as an integral of a continuous finite line of charge (see Eq.~\ref{intL0}).
For $m\geq0$, this definition generalizes straightforwardly by applying the ladder definition \eqref{pd+L} $m$ times: 
 \begin{align}
 L_n^m=&(-)^mR^{m+1}\partial_+^m \int_0^1\frac{ v^{n+m}\d v }{\sqrt{\rho^2+(z-Rv)^2}},\label{intLnm} 
 \end{align} 
Equivalently we could note that the integrand in \eqref{intLnm} is equivalent to
$v^{n+m} S_0^0(x,y,z-Rv)$, so by applying \eqref{pd+S} we can also write 
\begin{align}
	L_n^m=\int_{0}^1v^{n+m} S_m^m(x,y,z-Rv)\d v,
\end{align}
or more explicitly 
\begin{align}
	L_n^m=(2m-1)!!R^{m+1}\rho^me^{im\phi}\int_0^1\frac{ v^{n+m}\d v }{(\rho^2+(z-Rv)^2)^{m+1/2}},\label{intLnm2} 
\end{align}
which is a finite continuous line of multipoles of order $m$ with line density $v^{n+m}$.
In comparison, PSSHs have a similar integral representation with line density $(1-v^2)^{m/2}P_n^m(v)$ \cite{havelock1952moment,miloh1974ultimate}: 
\begin{align}
Q_n^m(\xi)P_n^m(\eta)e^{im\phi}
=&(-)^m(2m-1)!!R^{m+1}\rho^me^{im\phi}\nonumber\\
&\times\int_{-1}^1\frac{(1-v^2)^{m/2}P_n^m(v)\d v}{(\rho^2+(z-Rv)^2)^{m+1/2} }. \label{intQP}
\end{align}

\subsection{Series of multipoles}
The multipole series definition Eq.~\ref{LvsS0} can also be generalized using \eqref{pd+L} and \eqref{pd+S}, to obtain 
\begin{align}
L_n^m=\sum_{k=m}^{\infty} \frac{S_k^m}{n+k+1}\label{LvsS} \quad r>R.
\end{align}

Using this series, $L_n^m$ could in principle be defined for complex $n$ and integer $m$, as long as $n+m\neq-1,-2,-3...$.   Similarly the Legendre functions $Q_n^m$ are also finite only for $n\geq-m$. We will discuss a possible extension to $n<-m$ in Sec. \ref{sec L-n}.

\subsection{Finite sum of multipoles of the $2^\mathrm{nd}$ kind}

As for the $m=0$ case (Eq.~\ref{LvsSQ0}), the multipole series has an analytic continuation expressible in terms of SSHs of the 2$^{\rm nd}$ kind:
\begin{align}
L_n^m= \tilde{S}_n^{m} - \sum_{k=-m}^{n} {{n+m}\choose{k+m}}\tilde{S}_k^{m\prime}, \qquad n\geq-m.
\label{LvsSQ}
\end{align}
$\tilde{S}_n^m$ is singular on the entire $z$-axis while Eq. \eqref{LvsSQ} is singular only on the line segment $0\leq z\leq R$ due to cancellations between the functions. 

Eq.~\eqref{LvsSQ} contains $Q_k^m(u)$ for all $k\geq-m$. For $k\geq m$ they contain a logarithmic term, while for $k<m$ they are simply rational functions. Explicit expressions are tabulated in Ref.~\cite{suschowk1959explicit} (differing by $(-)^m$) for $k\geq0$ only, and we present some expressions for $k<0$ in appendix  \ref{secQnmneg}. For $k<m$, $Q_k^m(u)$ and $Q_{-k-1}^m(u)$ make up the two linearly independent solutions of the Legendre differential equation, while $P_{k<m}^m=0$.

\section{Generalization of logopoles to $m\geq0$: Properties}
\label{secLogopoles2}
We will now show how the definition for $m\geq0$ in the previous section translates to other properties of logopoles.

\subsection{Other differential operators}
Logopoles share similarities with SSHs regarding differential operators $\pd_z$ and $r\pd_r$, which both conserve harmonicity. The effect of these on $\tilde{S}_n^m$ are:
\begin{align}
R\pd_z \tilde{S}_n^m&=(n+m)\tilde{S}_{n-1}^m, \label{SQ dif} \\
r\pd_r \tilde{S}_n^m&=n\tilde{S}_n^m. 
\end{align}
Their effect on $L_n^m$ is similar but with an inhomogeneous part:
\begin{align}
R\pd_z L_n^m=&(n+m)L_{n-1}^m-S_m^{m\prime}, \label{dzL} \\
	r\pd_r    L_n^m=&nL_n^m-S_m^{m\prime}. \label{rdrL}
\end{align}
In contrast, the operators applied to the PSSHs result in infinite series:
\begin{align}
R\pd_z& ~ Q_n^m(\xi)P_n^{-m}(\eta)   \nonumber\\
&=-\sum_{\substack{k=n+1\\k+n\text{ odd}}}^\infty (2k+1) Q_k^m(\xi)P_k^{-m}(\eta) \label{pdzQP}\\
r\pd_r& ~ Q_n^m(\xi)P_n^{-m}(\eta)\nonumber\\
&=-\sum_{\substack{k=n\\k+n\text{ even}}}^\infty (2k+1-\delta_{nk}n) Q_k^m(\xi)P_k^{-m}(\eta) \label{rdrQP}\\
R\partial_+& ~Q_n^m(\xi)P_n^{-m}(\eta)e^{im\phi}\nonumber\\ 
& =-\sum_{\substack{k=n+1\\k+n\text{ odd}}}^\infty (2k+1) Q_k^{m+1}(\xi)P_k^{-m-1}(\eta)e^{i(m+1)\phi}.  \label{pd+QP}
\end{align}
\eqref{pdzQP} and \eqref{pd+QP} are derived in Ref.~\cite{matcha1971prolate}, while \eqref{rdrQP} is presented without proof. The Legendre functions of negative order $P_n^{-m}$ are here used only to simplify prefactors - see appendix \ref{AppLegendre} for their definitions.

\subsection{Recurrence relations} \label{sec rec}
Logopoles also obey similar recurrence relations to the SSHs. 
The recurrence relation on $n$ presented in Ref.~\cite{majic2019logopoles} generalizes to:
\begin{align}
&(n-m+1)L_{n+1}^m-\hat{z}(2n+1)L_n^m +\hat{r}^2(n+m)L_{n-1}^m \nonumber\\
&= (\hat{r}')^{1-m}P_m^m(\cos\theta')e^{im\phi}, \qquad\qquad\qquad n\geq m\label{Lrec}
\end{align}
which is similar to the recurrence for SSHs with an inhomogeneous term. It can be proved by substituting an explicit expression for the logopoles, for example Eq. \eqref{LvsSQ}.
In Sec. \ref{stable} this recursion is used to stably compute logopoles of high degree in the region $r<R$.

A recurrence relation on $m$ can be found by substituting the Legendre function recurrence
$2m P_n^m=\tan\theta(P_n^{m+1} + (n - m + 1) (m + n)P_n^{m-1}) $ into \eqref{LvsS}:
\begin{align}
L_n^{m+1}=&2m\cot\theta e^{i\phi} L_n^m  - [(n - m + 1) (m + n)L_n^{m-1} \nonumber\\
&+ S_m^{m-1\prime} - (n-m+1)S_{m-1}^{m-1\prime}]e^{2i\phi}. \label{Lrecm}
\end{align}
Unfortunately this is numerically unstable in the direction of increasing $m$ in all space, in particular near the $z$ axis.

For varying $m$ and $n$, there are multiple possible recurrence relations. For example by substituting the recurrence $\sin\theta P_n^{m+1}= (n+m+1)\cos\theta P_{n}^m -(n-m+1) P_{n+1}^m$ into \eqref{LvsS}:
\begin{align}
L_n^{m+1}=\frac{e^{i\phi}}{\sin\theta} \big[(n+m)\hat r L_{n-1}^m& -(n-m)\cos\theta L_n^m
\nonumber\\
&\qquad+(\cos\theta-\hat r)S_m^{m\prime}\big].\label{Lrecnm}
\end{align}
Again this is numerically unstable in the direction of increasing $m$, particularly near the $z$ axis.

\subsection{Explicit expressions of logopoles}
\label{sec explicit}

\begin{figure}
\includegraphics[scale=.68]{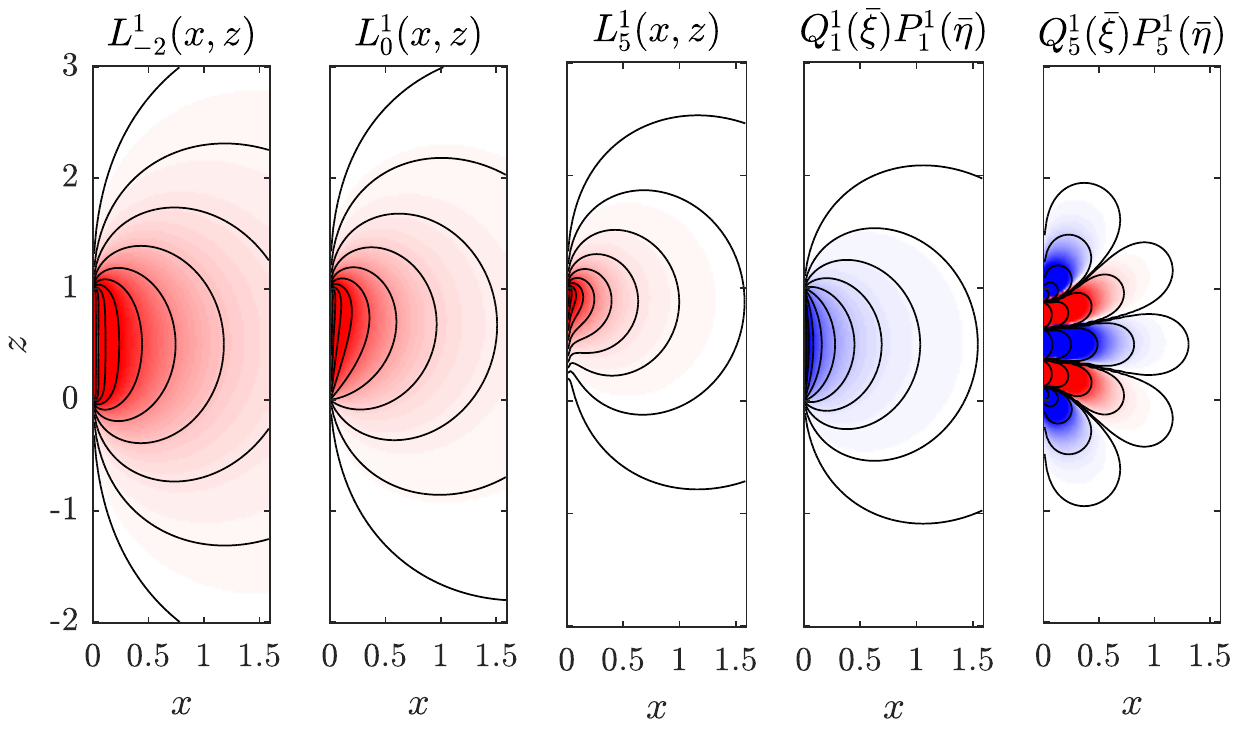}\\
\includegraphics[scale=.68]{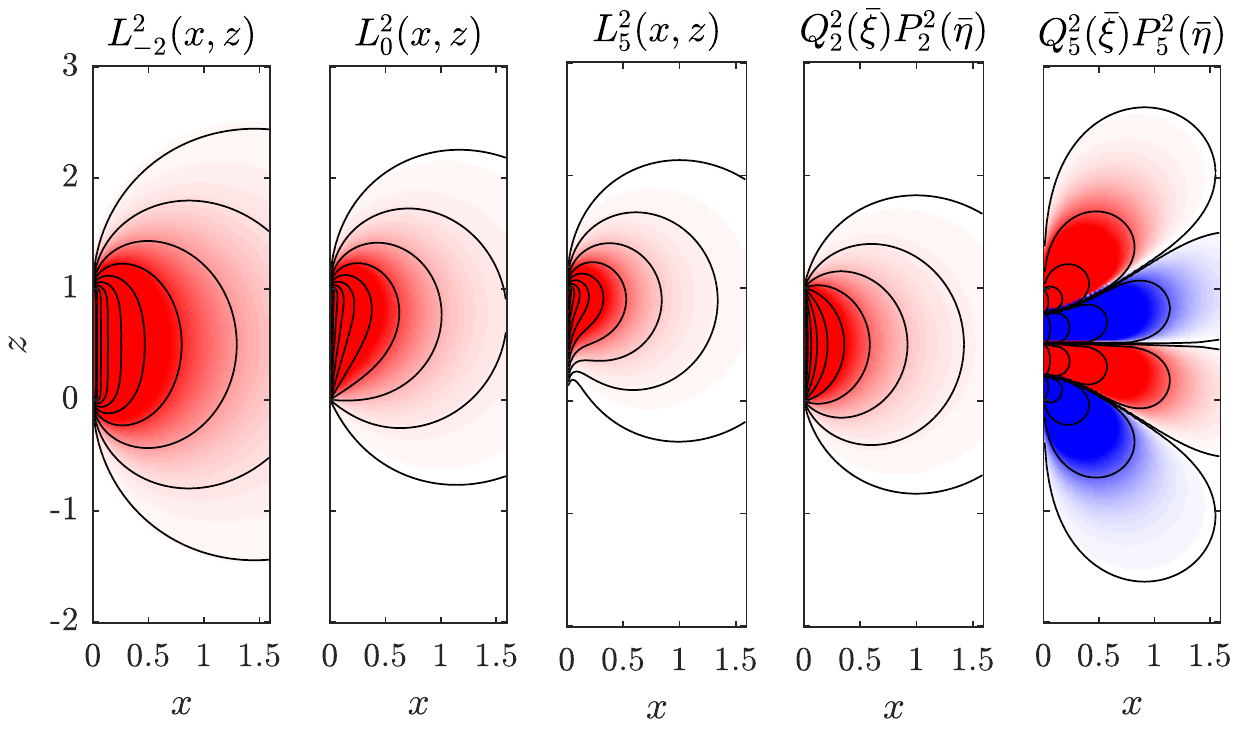}
\caption{Intensity plots of a representative selection of low degree logopoles and PSSHs for orders $m=1,2$. For better visualization, the functions have been transformed by scaling and taking the arcsinh, which is similar to plotting on a log scale but allows for negative values. Red is positive, white is zero, and blue negative. The PSSHs have been translated up the $z$-axis for a closer visual comparison. $R=1$. The logopoles and $Q_n^m$ functions were computed via forward recurrence which has no visible error for these low orders.}
\label{logoplots}
\end{figure}

Logopoles of low degree and order may be calculated for example via the finite sum of $\tilde{S}_k^m$ and $\tilde{S}_k^{m\prime}$ in Eq. \eqref{LvsSQ}. However, since this sum suffers numerically from catastrophic cancellation, it is best to use offset spheroidal coordinates $\xibar,\etabar$, where expressions do not explicitly involve subtraction of similar terms near the $z$ axis. Some low orders for $m=1,2$ are:
\begin{align}
L_{-1}^1=&~\tilde{S}_{-1}^1-\tilde{S}_{-1}^{1\prime}=R\frac{\cos\theta-\cos\theta'}{\rho}e^{i\phi}\label{L_1}\\
=&~\frac{4 \sqrt{1-\bar\eta ^2} \bar\xi }{\sqrt{\bar\xi ^2-1} \left(\bar\xi ^2-\bar\eta ^2\right)}e^{i\phi}\\
L_0^1=&~\frac{2}{\xibar-\etabar}\sqrt{\frac{1-\etabar^2}{\xibar^2-1}}e^{i\phi} \\
L_1^1=&~L_0^1+Q_1^1(\bar{\xi})P_1^1(\bar{\eta})e^{i\phi} \\
L_{-2}^2=&~8\frac{\bar\eta ^2-1}{\bar\xi ^2-1 }\frac{ 3 \bar\eta ^2 \bar\xi - \left(\bar\eta ^2-1\right) \bar\xi ^3 -3 \bar\xi ^5}{\left(\bar\xi ^2-\bar\eta ^2\right)^3}e^{i2\phi}\\
L_0^2=&~4\frac{2\xibar^2-1+\xibar\etabar}{(\xibar-\etabar)^3}\frac{1-\etabar^2}{\xibar^2-1}e^{i2\phi}\\
L_1^2=&-2\frac{3\xibar^2-2+\etabar^2}{(\xibar-\etabar)^3}\frac{1-\etabar^2}{\xibar^2-1} e^{i2\phi}.
\end{align}
Fig.~\ref{logoplots} plots a representative selection of logopoles and PSSHs.  For $n=-m$, these expressions coincide with \eqref{L_m stable}.

\subsection{Logopoles expanded as a series of offset multipoles}

In Ref. \cite{majic2019logopoles} logopoles were used to express the potential of a point charge near a dielectric sphere as a series of line images. This essentially required a multipole series for logopoles centred at O'. For dipolar or higher order multipole sources near a sphere, the generalization of this series is needed: 
\begin{align}
L_n^m=\sum_{k=m}^\infty (-)^{k+m}\frac{(n+m)!(k-m)!}{(n+k+1)!}S_k^{m\prime}  \qquad (r'>R)\label{LvsSp}.
\end{align}

\subsection{Separated form of logopoles} \label{sec separated}

Here we generalize the expression of logopoles in terms of their logarithmic and and non-logarithmic parts (Eq.~B3 in Ref.~\cite{majic2019logopoles}). Starting from the finite sum \eqref{LvsSQ}, we write $Q_n^m=P_n^mQ_0^0-W_{n-1}^m$, and use the translation relation for regular spherical harmonics \cite{hobson1944theory}:
\begin{align}
\sum_{k=m}^n\binom{n+m}{k+m} \hat r'^kP_k^m(u')=\hat r^nP_n^m(u),
\end{align} 
(note that $P_n^m=0$ for $-m<n<m$ and $m>0$) to get
\begin{align}
L_n^m=&\bigg(\hat r^n[P_n^m(u)L_0-W_{n-1}^m(u)] \nonumber\\
&+ \sum_{k=-m}^n\binom{n+m}{k+m}\hat r'^kW_{k-1}^m(u')\bigg)e^{im\phi}. \label{L isolated}
\end{align}
For $m=0$, $W_{n-1}^m(x)$ are polynomial, so the singularity on $0<z<R$ is entirely contained within $L_0$. 
For $m>0$ however, note that $W_{n-1}^m$ contain $(1-u^2)^{-m/2}$ which are singular on the $z$ axis; see appendix ~\ref{AppLegendre}.

\subsection{Relationships between logopoles and offset spheroidal harmonics} \label{secLvsQP}

The expression (\ref{QPbarvsL0}) of offset spheroidal harmonics as a finite sum of logopoles  can be generalized to
\begin{align}
Q_n^m(\xibar)&P_n^m(\etabar)e^{im\phi} = \frac{(n+m)!}{(n-m)!}\sum_{p=0}^n\frac{(-)^{p+n}(n+p)!}{2~p!(p+m)!(n-p)!} L_p^m. \label{QPvsL}
\end{align}
which may be derived from the integral representations, expanding the Legendre functions in terms of $(v+1)^n$:
\begin{align}
P_n^m&(v)(1-v^2)^{m/2}\nonumber\\
&= \frac{(n+m)!}{(n-m)!}\sum_{p=0}^n\frac{(-)^{p+n+m}(n+p)!}{2^p p!(p+m)!(n-p)!} (v+1)^{p+m}. \label{Pvsv+1}
\end{align}
and shifting $v\rightarrow 2v-1$.

But the inverse relationship expressing logopoles in terms of PSSHs (Eq.~\ref{LvsQPbar0}) cannot be simply generalized to $m>0$, and requires an infinite series. This can be obtained by substituting the series for $S_k^m$ in terms of $Q_p^m(\bar{\xi})P_p^m(\bar{\eta})$ from Ref.~\cite{majic2017super} into the multipole series for $L_n^m$:
\begin{align}
	L_n^m=&\sum_{k=m}^\infty\frac{S_k^m}{n+k+1}\nonumber\\
  	     =&\sum_{p=m}^\infty\bigg[\sum_{k=m}^p\frac{2p+1}{n+k+1}\frac{2(p+k)!(-)^{p+k+m}}{(p-k)!k!(k-m)!}\frac{(p-m)!}{(p+m)!}\bigg]\nonumber\\
  	     &\hspace{3cm}\times Q_p^m(\bar{\xi})P_p^m(\bar{\eta})e^{im\phi}. \label{LvsQPdouble}
\end{align}  
We may define the expansion coefficients $\beta_{np}^m$ as the sum over $k$:     
\begin{align}
  	      L_n^m= \sum_{p=m}^\infty\beta_{np}^mQ_p^m(\bar{\xi})P_p^m(\bar{\eta})e^{im\phi}.\label{LvsQPcoefs1}	     
\end{align}
For $m=0$ only, the coefficients vanish if $p>n$. Unfortunately for $m>0$ the sum in \eqref{LvsQPdouble} for $\beta_{np}^m$ is numerically unstable and produces large errors for $p\gtrsim20$. See appendix \ref{sec beta} for more details.

\subsection{Relationships between logopoles and centred spheroidal harmonics} \label{secLvsQPcentred}

Spheroidal harmonics centered about the origin with foci at $z=-R,R$ can be expressed in terms of geometrically transformed logopoles, shifted and flipped in the $z$ direction:
\begin{align}
L_n^m(z\rightarrow R-z)=&(-)^mR^{m+1}\pd_+^m\int_0^1\frac{ (1-v)^{n+m}\d v }{\sqrt{\rho^2+(z-Rv)^2}}, \label{L R-z}\\
L_n^m(z\rightarrow R+z)=&(-)^mR^{m+1}\pd_+^m\int_{-1}^0\frac{ (1+v)^{n+m}\d v }{\sqrt{\rho^2+(z-Rv)^2}}. \label{L R+z}
\end{align}
The integral form of $Q_n^m(\xi)P_n^m(\eta)$, \eqref{intQP}, contains the integrand $(1-v^2)^{m/2}P_n^{m}(v)$. To expand this in powers of $v+1$ and $v-1$ we take the expansion of $P_n(v)$ in terms of powers of $(v-1)$, and integrate $m$ times by $\int_v^1\d v$ (using 8.752.2 in Ref.~\cite{tables2014}) to give
\begin{align}
(1-v^2)^{m/2}P_n^{-m}(v)=\sum_{p=0}^n\frac{(-)^m(n+p)!(1-v)^{p+m}}{2^p p!(p+m)!(n-p)!}.
\end{align}
Comparing this with the line integral expressions for $Q_n^m(\xi)P_n^m(\eta)$ (Eq. \ref{intQP}), $L_p^m(z\rightarrow z-R)$ (Eq. \ref{L R-z}), and $L_p^m(z\rightarrow R+z)$ (Eq. \ref{L R+z}), we deduce 
\begin{align}
Q_n^m(\xi)&P_n^{-m}(\eta)e^{im\phi} = \sum_{p=0}^n\frac{(-)^p(n+p)!}{2^{p+1}p!(p+m)!(n-p)!} \nonumber\\&\times[L_p^m(z\rightarrow R-z)+(-)^{n+m}L_p^m(z\rightarrow R+z)]. \label{QPvsLsa}
\end{align}


\subsection{Orthogonality and completeness} \label{sec orthogonality}

Logopoles are positive functions (for $\phi=0$) which means they cannot be orthogonal with respect to the $L^2$ inner product with a positive weight function, because the inner product of any two logopoles will always be greater than zero.
And unlike the SSHs or the PSSHs, logopoles do not form a complete basis for potentials which decay as $1/r$ towards infinity, despite the fact that they are a finite linear combination of PSSHs.
As a specific example we attempt to expand Green's function $1/|\b r-\b r_0|$ for some point $\b r_0$.
An attempt can be made starting with the expansion in terms of $Q_n(\bar\xi)P_n(\bar\eta)$:
\begin{align}
\frac{R}{|\b r-\b r_0|} &=\sum_{m=-\infty}^\infty\sum_{n=|m|}^{\infty}2(2n+1)Q_n^m(\bar\xi)P_n^{-m}(\bar\eta)\nonumber\\	&\hspace{1cm}\times P_n^m(\bar\xi_0)P_n^{-m}(\bar\eta_0) \qquad \bar{\xi}>\bar{\xi}_0
\end{align} 
then applying the expansion of spheroidal harmonics in terms of logopoles (Eq. \ref{QPvsL})
and rearranging the order of summation:
\begin{align}
\frac{R}{|\b r-\b r_0|} 
&=\sum_{m=-\infty}^\infty\sum_{k=0}^\infty\sum_{n=\max(|m|,k)}^\infty  \!\!\!\!(2n+1)P_n^m(\bar\xi_0)P_n^{-m}(\bar\eta_0)\nonumber\\
&\hspace{1cm}\times\frac{(-)^{n+k+m}(n+k)!}{k!(k+m)!(n-k)!}L_k^m(\b r)
\end{align}
but the coefficient of $L_k(\b r)$ is given by a series over $n$ which diverges for any $\b r_0$ - the coefficients increase with the exponential factor in $P_n(\bar\xi_0)\rightarrow\sim \xi_0^n$ for $\xi_0>1$, while $P_n(\bar\eta_0)$ behaves sinusoidally with $n$ \cite{NIST:DLMF}, and for $\xi_0=1$ ($\b r_0$ on the line segment) the coefficients increase with the polynomial $(n+k)!/(n-k)!$. 
This is related to the fact that the monomial basis $v^n$ is incomplete in that it cannot expand the delta function $\delta(v-z_0)$ which is the charge distribution of the Green for $\b r_0$ on the line segment. Instead, the Legendre basis $P_n(v)$ $can$ be used to expand the delta function, as  $\delta(v-z_0)=\sum_{n=0}^\infty(2n+1)P_n(z_0)P_n(v)$.


\section{New relationship between spheroidal and spherical harmonics of the 2$^\mathrm{nd}$ kind}
\label{secNewRelationship}

We now present the generalization of the expansion of PSSHs in terms of SSHs of the 2$^{\rm nd}$ kind (Eq.~\ref{newrel0}):
\begin{align}
&Q_n^m(\xi)P_n^{-m}(\eta)=\sum_{k=0}^n\frac{(n+k)!}{2k!(k+m)!(n-k)!}\nonumber\\
&\times\left[(-)^{n+k+m}\left(\frac{r''}{2R}\right)^kQ_k^m(u'')-\left(\frac{r'}{2R}\right)^kQ_k^m(u')\right]. \label{QPvsSQpandSQppm} 
\end{align}
As for $m=0$, the linear combination of offset SSHs of the 2$^{\rm nd}$ kind ensures that their singularities exactly cancel for $|z|>R$, so the resulting sum is only singular on the segment from O' to O''.
Eq. \eqref{QPvsSQpandSQppm} can be proved by substituting the expansion of offset spheroidal harmonics in terms of logopoles (Eq. \ref{QPvsL}), and expanding the logopoles in terms of $\tilde{S}_n^m$ and $\tilde{S}_n^{m\prime}$. A double sum arises which can be simplified using the same identity as used for the $m=0$ case. \\
Similar to the finite sum for logopoles Eq. \eqref{LvsSQ}, the sum here involves $Q_k^m(u)$ with $k<m$ which are linearly independent from $Q_k^m(u)$ and $P_k^m(u)$ with $k\geq m$, and do not contain a logarithmic term (see appendix \ref{secQnmneg}). \\

It is insightful to look at the logarithmic part of Eq.~\eqref{QPvsSQpandSQppm}.
Using the identity \eqref{QPQ0R}, $Q_n^m(\xi)=Q_0^0(\xi)P_n^m(\xi)-W_{n-1}^m(\xi)$, we can break down Eq.~\eqref{QPvsSQpandSQppm} and examine the $Q_0^0(\xi)$ part. 
The regular spheroidal harmonics $P_n^m(\xi)P_n^m(\eta)$ can be expanded in terms of offset solid spherical harmonics:
\begin{align}
P_n^m&(\xi)P_n^{-m}(\eta)=\!\sum_{k=m}^n\frac{(-)^{n+k+m}(n+k)!}{k!(k+m)!(n-k)!}\!\left(\frac{r''}{2R}\right)^k\!\!P_k^m(u'') \label{PPnmvsPnm''}
\end{align}
which is proved in Ref.~\cite{majic2017super} in an offset frame. $P_n^m(\xi)P_n^m(\eta)$ can also be expanded in terms of $S_k^{m\prime}$ with similar coefficients by reflecting Eq.~\eqref{PPnmvsPnm''} about $z=0$ and using the parity of the Legendre polynomials. 
Multiplying Eq.~\eqref{PPnmvsPnm''} by $Q_0^0(u'')$, and the corresponding $S_k^{m\prime}$ expansion by $Q_0^0(u')$, subtracting these equations and noting the relation $Q_0^0(\xi)=[Q_0^0(u'')-Q_0^0(u')]/2$ leads to
\begin{align}
P_n^m(&\xi)Q_0^0(\xi)P_n^{-m}(\eta) = 
\sum_{k=m}^n\frac{(n+k)!}{2k!(k+m)!(n-k)!}
\nonumber\\
&\times\Bigg[(-)^{n+k+m}\left(\frac{r''}{2R}\right)^kP_k^m(u'')Q_0(u'')
\nonumber\\
&\qquad\qquad\qquad-\left(\frac{r'}{2R}\right)^kP_k^m(u')Q_0(u')\Bigg],\label{singular part eq}
\end{align}
which provides a simple proof of the equality of the logarithmic part of Eq. \eqref{QPvsSQpandSQppm}.

\section{Integral representation of SSHs of the 2$^{\rm nd}$ kind} \label{secLineInts}

In Ref.~\cite{majic2019logopoles}, the regular solid spherical harmonics of the second kind for $m=0$ were shown to be equivalent to a line source distribution on the entire $z$-axis, regularized by a sum of multipoles from sources at infinity, of infinite magnitude. By applying $\tilde{S}_n^m=(-R)^m\pd_+^m \tilde{S}_{n+m}$ to the $m=0$ case, we get, for $n\geq-m$:
\begin{align}
	\tilde{S}_n^m=\lim_{\mu\rightarrow\infty} \bigg\{&\frac{(2m-1)!!}{2}\rho^m\int_{-\mu}^\mu\frac{R^{m+1}\text{sign}(v)v^{n+m}\d v}{(\rho^2+(z-Rv)^2)^{m+1/2}} \nonumber\\
	& -\sum_{\substack{k=m\\n-k\text{ odd}}}^{n-1}\frac{\mu^{n-k}}{n-k}\hat r^{k}P_{k}^m(u) \bigg\}e^{im\phi}.\label{intSQ}
\end{align}
which is the potential of sources on the $z$ axis and at $r=\infty$. For $n\leq m$ the sum over $k$ vanishes and the integral is finite so no regularization is needed. The radius $R$ only appears here since it is included as a scaling factor in the definition of $\tilde{S}_n^m$. Eq. \eqref{intSQ} has been checked in Mathematica for small integer $n$ and $m$. Comparison with the integral expression for logopoles, Eq. \eqref{intLnm} shows that the source distributions of logopoles are finite truncations (or regularizations) of that for SSHs of the 2$^{\rm nd}$ kind.\\

Eq. \eqref{intSQ} can also be rewritten as to regularize the integrand itself, avoiding the need for limits:
\begin{align}
	\tilde{S}_n^m=\frac{1}{2}\int_{-\infty}^{\infty} \bigg(&\text{sign}(v)v^{n+m} S_m^m(x,y,z-Rv)   \nonumber\\
	& -\sum_{\substack{k=m\\n-k\text{ odd}}}^{n-1}v^{n-k-1}\hat S_k^m(x,y,z) \bigg) \d v .\label{intSQ2}
\end{align}

For completeness we present the analogous formula for the \textit{exterior} SSHs of the 2$^{\rm nd}$ kind, which can be derived by applying the Kelvin transform $\tilde{S}_n^m\rightarrow \tilde{S}_n^m(r\rightarrow1/r)/r$:
\begin{align}
	\hat r^{-n-1}Q_{n}^m(u)=&\lim_{\mu\rightarrow 0} \bigg\{\frac{(2m-1)!!}{2} \rho^m \nonumber\\
	&\times\bigg(\int_{-\infty}^{-\mu}+\int_{\mu}^\infty\bigg)\frac{R^{m+1}v^{-n-1+m}\d v}{(\rho^2+(z-Rv)^2)^{m+1/2}} \nonumber\\
	& -\sum_{\substack{k=m\\n-k\text{ odd}}}^{n-1}\frac{\mu^{k-n}}{n-k}\hat r^{-k-1}P_{k}^m(u) \bigg\}.\label{intSQr-n-1}
\end{align}
for $n\geq-m$. This line source distribution diverges at the origin, and is regularized by a finite sum of multipoles of infinite strength. Eq. \eqref{intSQr-n-1} can also be rewritten as to regularize the integrand itself:
\begin{align}
	\hat r^{-n-1}Q_{n}^m(u)e^{im\phi}= \frac{1}{2}\int_{-\infty}^{\infty}& \bigg(v^{-n-1+m}S_m^m(x,y,z-Rv)\nonumber\\
	&-\sum_{\substack{k=m\\n-k\text{ odd}}}^{n-1}v^{k-n-1}S_k^m(x,y,z)\bigg)\d v. \label{intSQr-n-12}
\end{align}


\section{Further generalization to negative degrees and orders}
\label{negative}

The sum \eqref{LvsSQ} provides a means to define logopoles of negative degree when $n\geq -m$. We now discuss their properties in  detail.

\subsection{Lowest degree: $n=-m$} \label{sec L_m}

\subsubsection{Definition}
For a given $m$, the lowest integer $n$ for which $L_n^m$ is finite is $n=-m$. The integral expression Eq. \eqref{intLnm} shows that $L_{-m}^m$ is the potential of a uniform distribution of transverse $2^m$ multipoles on the line segment $0<z<R$. 
 The explicit formulae from the analytic continuation Eq. \eqref{LvsSQ} is
 \begin{align}
 L_{-m}^m=&\tilde S_{-m}^m-\tilde S_{-m}^{m\prime}\nonumber\\
 =&\frac{e^{im\phi}}{\hat\rho^m}\big[\sin^m\theta Q_{-m}^m(\cos\theta)-\sin^m\theta' Q_{-m}^m(\cos\theta')\big]. \label{L_m vs Q}
 \end{align}
%
%
 $(1-u^2)^{m/2}\theta Q_{-m}^m(u)$ is an odd polynomial in $u$ (see appendix \ref{secQnmneg}):
 \begin{align}
 Q_{-m}^m(u)=\frac{(2m-1)!!}{(1-u^2)^{m/2}}\sum_{k=0}^{m-1}\frac{(-)^{m+k}}{2k+1}\binom{m-1}{k}u^{2k+1}. \label{Q-mm}
 \end{align} 
%
 While the explicit expression Eq. \eqref{L_m vs Q} is simple, it is numerically unstable near the $z$-axis for $z<0$ and $z>R$, showing catastrophic cancellation particularly for higher $m$, since it is expressed as the difference of two functions which individually diverge toward the $z$-axis (as $\rho^{-m}$), but their subtraction goes to zero (as $\rho^m$). See Sec. \ref{SecStable-m} for a numerically stable expression.
 
\subsubsection{Angular integral expression for $L_{-m}^m$}
From Eq.~\eqref{Q-mm}, we can deduce that the Legendre functions of the second kind may instead be represented as an integral:
\begin{align}
Q_{-m}^m(\cos\theta)=(-)^m\frac{(2m-1)!!}{\sin^m\theta}\int_\theta^{\pi/2}\sin^{2m-1}\bar\theta~\d \bar\theta. \label{Q-mm int}
\end{align}
Then from Eq. \eqref{L_m vs Q} we have
\begin{align}
L_{-m}^m=(-)^m(2m-1)!!\frac{R^m}{\rho^m}\int_{\theta}^{\theta'}\sin^{2m-1}\bar\theta~\d\bar\theta ~e^{im\phi}.
\end{align}
This form is similar to that used in Ref. \cite{curtright2016charged} to describe uniformly charged line segments in $2m+3$ dimensions where the electrostatic force decays as $1/r^{2m+1}$.  This aligns with the interpretation of $L_{-m}^m$ as a distribution of transverse $2^m$ multipoles on the line segment which share this $1/r^{2m+1}$ decay factor.

\subsubsection{Recurrence relation for $L_{-m}^m$} \label{Sec rec-m}
Logopoles $L_{-m}^m$ obey a first order recurrence relation for increasing $m$:
\begin{align}
L_{-m-1}^{m+1}=&2\frac{m}{\rho}L_{-m}^m e^{i\phi} \label{Lrec_m}\nonumber\\& +(2m-1)!!\rho^{m-1}\bigg(\frac{u}{r^{2m}}-\frac{u'}{r'^{2m}}\bigg)e^{i(m+1)\phi},
\end{align}
which can be derived from Eq. 2.263.3 in Ref.~\cite{tables2014}.
  Eq. \eqref{Lrec_m} is stable in the forward direction (increasing $m$) at least for $\rho>0$, $0<z<R$, due to the fact that in this region, both $L_{-m}^m$ and the inhomogeneous term have positive sign, so the two terms are added with no loss of precision. This stability is confirmed by numerical tests.

\subsection{Logopoles for $n<0$,~$~m=0$} \label{sec L-n}
One can also generalize further to $n<-m$, with modifications. For simplicity we consider $m=0$. For
for $n=-1,-2,-3...$, the series definition \eqref{LvsS} has a singular term, which can be avoided if we redefine logopoles for $n<0$ to start the series from $k=n$:
\begin{align}
L_{-n}\equiv\sum_{k=n}^\infty\frac{S_k}{k-n+1}. \label{L-n}
\end{align} 
These logopoles can be expressed as convergent integral by expanding the integrand of \eqref{intL0} in terms of spherical harmonics and truncating the series as done in \eqref{L-n}:
\begin{align}
	L_{-n} = \int_0^1  v^{-n}\bigg[S_0(x,y,z-Rv) - \sum_{k=0}^{n-1}v^k S_k(x,y,z)\bigg]\d v 
\end{align}
which shows that $L_{-n}$ has a structure of a line charge on the segment $0<z\leq R$ with distribution $z^{-n}$, regularized by multipoles of degree $k=0,1,...,n-1$ of infinite strength. This is similar to the charge distribution for the SSHs of the 2$^{\rm nd}$ kind in Eq. \eqref{intSQr-n-12}.

In fact, $L_{-1}$ is used as an approximate solution in Ref. \cite{ranvcic2006point} for the reflected potential of a point charge near a dielectric sphere of high relative dielectric constant $\epsilon$.  $L_{-1}$ has a finite line charge distribution of $1/z$, which closely matches the image line charge $\propto z^{-\epsilon/(\epsilon+1)}$ when $\epsilon$ is large. It has a simple closed form expression:
\begin{align}
L_{-1}=\frac{R}{r} \log\frac{2r}{r-Ru+r'}. \label{L-1}
\end{align}

\begin{figure}
	\includegraphics[scale=.66]{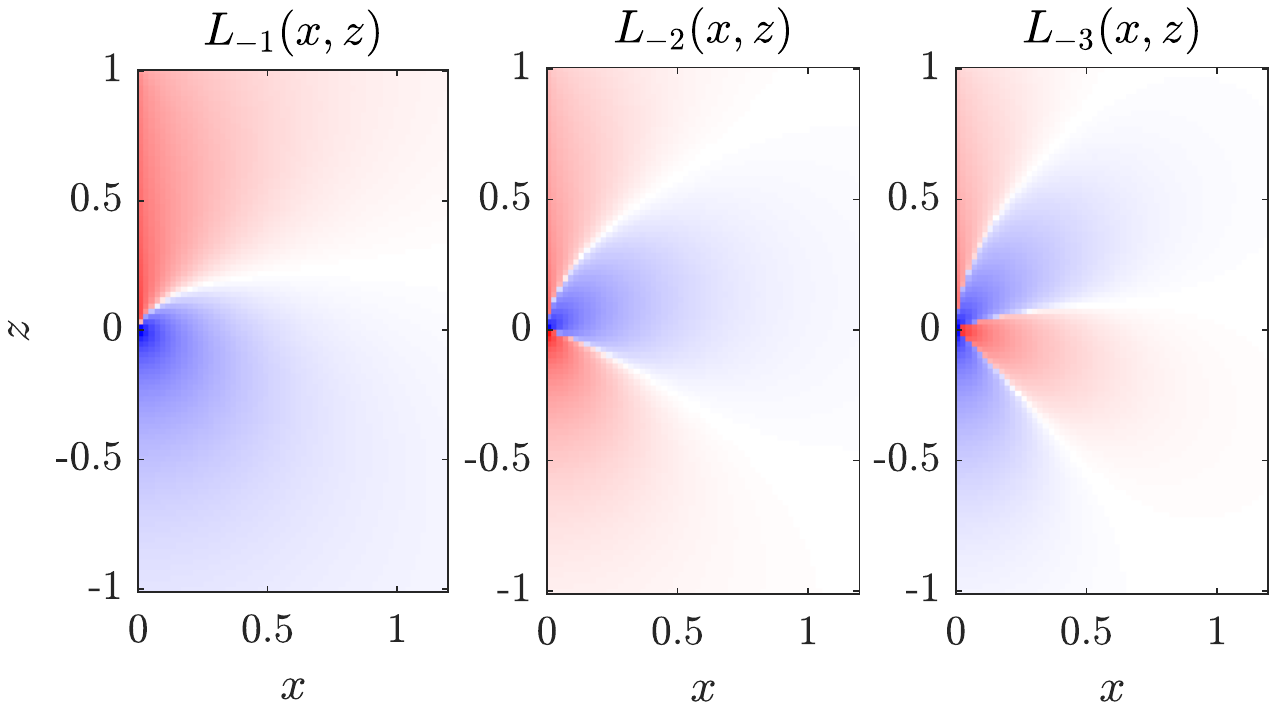}
	\caption{A representative sample of logopoles for $n<0$. The arcsinh of the functions is plotted for better visualization. Red is positive, blue negative, white zero. $R=1$. The functions are singular on the axial line segment $0
	\leq z\leq R$.} \label{logoplots_n}
\end{figure}

The recurrence properties of these functions are more complicated than for $n\geq 0$. By applying $\pd_z$ to \eqref{L-n} we find:
\begin{align}
R\pd_z L_{-n}= - nL_{-n-1} -\frac{R}{r'} + \sum_{k=0}^n S_k, \label{difL-n}
\end{align}
and by applying the recurrence for the Legendre polynomials we find:
\begin{align}
nL_{-n}=(2n+1)&\frac{z}{R}L_{-n-1} - (n+1)\frac{r^2}{R^2}L_{-n-2} \nonumber\\
&- \frac{r'}{R}+ \frac{r'^2}{R^2}\sum_{k=0}^nS_k + \frac{r^2}{R^2}S_{n+1}-S_n. \label{recL-n} 
\end{align}
This recurrence relation was used to find explicit formulae for the lowest degrees. These are expressed stably in terms of spherical harmonics and offset spheroidal coordinates as
\begin{align}
	L_{-1}=&S_0\log\frac{(\bar\xi+\bar\eta)^2}{\bar{\xi}^2-1} \\
	L_{-2}=&S_1\log\frac{(\bar\xi+\bar\eta)^2}{\bar{\xi}^2-1} -  4\frac{1-\bar\eta^2}{(\bar{\xi}+\bar{\eta})^3} \\
	L_{-3}=&S_2\log\frac{(\bar\xi+\bar\eta)^2}{\bar{\xi}^2-1} -  2(7 + \bar\eta^2 + 8\bar\eta\bar\xi)\frac{1- \bar\eta^2}{(\bar\xi+\bar\eta)^5} \\
	L_{-4}=&S_3\log\frac{(\bar\xi+\bar\eta)^2}{\bar{\xi}^2-1} - \frac{4}{3}\frac{1-\bar\eta^2}{(\bar\xi+\bar\eta)^7}\nonumber\\ &\times\big[37 - 2\bar\eta^2 + \bar\eta^4 + 9\bar\eta(7 + \bar\eta^2)\bar\xi + 9(5\bar\eta^2-1)\bar\xi^2\big],
\end{align}
which are plotted in Fig. \ref{logoplots_n}, clearly showing the multipoles of infinite strength.

\subsection{Logopoles for $m<0$} \label{sec m<0}
There is a unique generalization of logopoles to $m<0$, which can be realized by applying the relationships for Legendre functions of negative and positive order to the finite sum \eqref{LvsSQ}:
\begin{align}
	L_n^{-m}
	=&(-)^m\frac{(n-m)!}{(n+m)!}\bigg[\tilde{S}_n^m-\sum_{k=m}^{n}\binom{n+m}{k+m}\tilde{S}_k^{m\prime}\bigg].\label{Lm<0}
\end{align}
However, $L_n^{m<0}$ are singular on an unbounded region on the $z$-axis.

\section{Numerical computation of logopoles}
\label{SecStable}
Numerical computation of logopoles is unstable for some parameter ranges, which could hamper their applications to physical problems. Here we investigate this numerical instability  and present  methods of stable computation.

\subsection{Computation on the $z$-axis for $L_n^0$}
We first investigate the nature of the numerical cancellations/instability in the computation of logopoles, on the $z$-axis for $z>R$ and $m=0$. Starting from Eq.~\eqref{L isolated} for $u,u'=1$, the functions $W_{n-1}$ reduce to the harmonic numbers $H_n$:
\begin{align}
	L_n(\rho=0,z>R)=&\hat{z}^n\bigg(\log\frac{\hat{z}}{\hat{z}-1}-H_n\bigg) \nonumber\\
	& + \sum\limits_{k=0}^{n}\binom{n}{k}(\hat{z}-1)^k H_k \label{L zaxis1},
\end{align} 
which can be rearranged using the binomial expansion and an identity proved in Ref.~\cite{majic2019logopoles} to give
\begin{align}
	L_n(\rho=0,z>R)=&\hat{z}^n\bigg(\log\frac{\hat{z}}{\hat{z}-1} - \sum_{k=1}^{n} \frac{\hat{z}^{-k}}{k} \bigg). \label{L zaxis2}
\end{align}
 As $n$ increases, the log term and the sum in Eq. \eqref{L zaxis2} become increasingly similar in magnitude, especially for large $\hat z$, so their difference becomes too small to accurately express with floating point arithmetic. Finding a numerically stable closed form of logopoles for large $n$ reduces to finding a stable expression for Eq. \eqref{L zaxis2}. We can write the log term as a series to obtain
\begin{align}
L_n(\rho=0,z>R)=&\sum_{k=n+1}^\infty \frac{\hat{z}^{n-k}}{k}, \label{L zaxis3}
\end{align}
which is numerically stable, and equivalent to the multipole series \eqref{LvsS}. These numerical considerations apply to $\rho>0$ too, and one can always compute logopoles stably via Eq. \eqref{LvsS} (for $r>R$).

\subsection{Stable explicit expression for $L_{-m}^m$} \label{SecStable-m}
 To compute $L_{-m}^m$ stably, it is necessary to break up the primed and unprimed components in Eq.~\ref{L_m vs Q} and extract from their difference a factor that governs the behavior both near the $z$ axis for $z<0$ or $z>R$  where $L_{-m}^m\rightarrow\rho^m$, 
and also near the singularity where $L_{-m}^m$ diverges as $\rho^{-m}$. This dependence is expressed naturally in spheroidal coordinates as a factor of $\big(\frac{1-\bar\eta^2}{\bar\xi^2-1}\big)^{m/2}$, which has exactly this behaviour. 
In appendix \ref{stableproof} we rearrange Eq. \eqref{L_m vs Q} into a form where this is factored out explicitly: 
\begin{align}
L_{-m}^m=	&\frac{2^{m+1}(2m-1)!!}{\left(\bar\xi ^2-\bar\eta ^2\right)^{2m-1}}\left(\frac{1-\bar\eta^2}{\bar\xi^2-1}\right)^{m/2} e^{im\phi} \nonumber\\
&\times
\sum _{q=0}^{m-1} \frac{\bar\xi ^{2 q+1}}{2 q+1}  \binom{m-1}{q} \left(\bar\xi ^4-\bar\eta ^2\right)^{m-2 q-1} \nonumber\\ &\times \sum _{p=0}^q  (-)^{p+q}\binom{q}{p}\frac{ (2 m+2 p-2 q-3)!!}{(2 p-1)!! (2 m-2 q-3)!! }  \nonumber\\
&\quad\times\left(\bar\eta  \left(\bar\xi ^2-1\right)\right)^{2p}\left(\bar\xi ^2-\bar\eta ^2\right)^{2q-2p}. \label{L_m stable}
\end{align}
Eq. \eqref{L_m stable} appears complicated but is nevertheless stable in almost all space, particularly near the $z$ axis for $|z|>R$. The worst relative error occurs near the singularity for $|z|<R$, and increases only mildly with $m$, reaching $\approx 10^{-5}$ at $m=40$ which should be fine for practical applications. 
Alternatively, the recurrence \eqref{Lrec_m} can be used in this region $|z|<R$, where it is stable for increasing $m$.

\subsection{Stable implementation of recurrence relations} \label{stable}

Here we propose a stable method of computing the logopoles for a large range of indices $n,m$ in all regions of space. For this, we calculate $L_n^m$ via a combination of forward and backward recurrence on $n$, Eq. \eqref{Lrec}, depending on the region of space. 
Eq. \eqref{Lrec} is numerically stable in the forward direction for $r<R$, unstable otherwise. For example at the point $\rho=2R$, $z=0$, numerical errors become of order $~1$ at $n\approx40$.
We can look at this stability analytically by considering the characteristic polynomial - the recurrence is stable if the roots of the characteristic polynomial are all less than 1 in magnitude. Although the characteristic polynomial is \textit{inhomogeneous}, in the large $n$ limit, it becomes equivalent to that for the homogeneous recurrence because the homogeneous terms in Eq. \eqref{Lrec} grow as $\cO(n)$ relative to the
inhomogeneous term. The characteristic equation for the homogeneous part is
\begin{align}
(n-m+1)\lambda^2-(2n+1)\hat{z}\lambda+(n+m)\hat{r}^2=0. \label{chr eq}
\end{align}
In the limit $n\rightarrow\infty$, the roots are $\lambda=\hat{r}e^{\pm i\theta}$ so the recurrence is stable for $r<R$ and unstable for $r\geq R$. This explanation strictly only applies to the large $n$ limit, but for finite $n$ the same conclusion is drawn from numerical tests.

Conversely, in the backward direction, Eq. \eqref{Lrec} is stable for $r>R$ and unstable for $r\leq R$.
Numerical tests confirm these regions of stability unless $r'\rightarrow0$ or $m$ is large. 
Hence we propose to calculate $L_n^m$ via Eq. \eqref{Lrec} forwards for $r<R$ and backwards for $r>R$.\\

In terms of implementing the forward recursion over $n$, two initial values are needed, first, $L_{-m}^m$, calculated via Eq. \eqref{L_m stable} for $z<0,z>R$ and \eqref{L_m vs Q} for $0<z<R$ as per Sec \ref{sec L_m}, and $L_{-m+1}^m$, obtained from: 
\begin{align}
	L_{-m+1}^m=\hat zL_{-m}^m + (2m-3)!!\hat\rho^m\left[\hat{r}^{1-2m}-\hat{r}^{\prime 1-2m}\right],
\end{align}
which can be derived from Eq. 2.263.3 in Ref.~\cite{tables2014}.\\
A singular point occurs in the forward recurrence at $n=m-1$, for calculating $L_m^m$ from $L_{m-1}^m$ and $L_{m-2}^m$, due to the division of $n-m+1$. Instead to calculate $L_m^m$ we must use a single triangular recursion step which requires a logopole of degree $m-1$: (from Eq. 2.263.2 in Ref.~\cite{tables2014})
\begin{align}
	L_m^m=-(2m-3)!!\frac{\hat\rho^m}{\hat r'^{2m-1}} + \hat z L_{m-1}^m +\hat\rho L_{m-1}^{m-1}
\end{align}
and from there the forward recurrence Eq. \eqref{Lrec} can be continued starting with calculating $L_{m+1}^m$. 

For backward recursion, we use initial values $L_N$, $L_{N+1}$ for large $N$, which may be taken as approximate from the limit $N\rightarrow\infty$: 
\begin{align}
	L_{n\rightarrow\infty}^m\rightarrow \frac{S_m^{m\prime}}{n+m+1}.
\end{align} 
Even though this is approximate, the recursion corrects itself quickly as $n$ decreases. If the logopoles are required to be calculated for  $n\leq N'$, then $N$ should be taken large enough so that the recurrence corrects itself to within floating point precision by the stage it reaches $n=N'$. The logopoles will satisfy the correct recursion, but be scaled incorrectly. The correct scaling can be determined from the known expression for the lowest degree $L_{-m}^m$, given by $L_0=\text{atanh}(\bar{\xi})$ for $m=0$ or by Eq. \eqref{L_m stable} for $m>0$. 
This recursive scheme can compute logopoles accurately in all space for a reasonable range of parameters. For a given $m$, the error becomes significant for all values higher than some $n$. For $m=0,1,2$, the method seems stable for arbitrarily large $n$, at least 400. For $m=7$, error becomes order 1 at $n\approx100$, for $m=10$, at $n\approx25$, and for $m=20$, at $n\approx20$, with errors first appearing near $r=R$.

\section{Conclusion}

We have presented a natural generalization of logopoles to $m\geq0$, based on the application of the differential operator $\pd_x+i\pd_y$ and its effect on spherical harmonics. Using this generalization, most properties of logopoles for $m=0$ are retained; sharing many similarities with spherical harmonics of the second kind, in their recurrence relations and source distributions, while having the advantage of a bounded singularity.
This generalization involves the Legendre functions $Q_{n}^m(\cos\theta)$ for $n<m$, which are also found to  relate to the canonical spheroidal harmonics $Q_n^m(\xi)P_n^m(\eta)$, defined only for $n\geq m$. 
 The line source integral expressions have been generalized for logopoles, spheroidal harmonics and spherical harmonics of the second kind with radial factors $r^n$ and $r^{-n-1}$. For $m=0$, we have also considered a possible generalization of logopoles for $n<0$, whose source distributions are similar to those for spherical harmonics of the second kind with $r^{-n-1}$ radial dependence. Methods of stable computation of logopoles have been investigated, including forward and backward recursion and different sum and series expressions, which are ideal in different regions of space. 
 
This work completes the introduction of logopoles and sets the foundation for further investigations and applications; application of logopoles of order $m=1$ to the problem of an electrostatic dipole near a dielectric sphere will be presented in a separate publication.

\section{Data availability}
The data that support the findings of this study are available from the corresponding author upon request.

\begin{acknowledgments}
We want to thank an anonymous reviewer who made contributions and gave insights in terms of integral representations, and helped to check formulas.\\
MM acknowledges support though a Victoria University of Wellington doctoral scholarship.
ECLR acknowledges the support of the Royal Society Te Ap\=arangi (New Zealand) through a Marsden grant.
\end{acknowledgments}

\appendix


\section{Legendre functions} 

\subsection{Positive degree $n$}
\label{AppLegendre}

The associated Legendre functions of the first and second kinds of any order $m\geq0$ can be defined from the Legendre functions of order $m=0$ via
\begin{align}
P_n^m(x)&=|1-x^2|^{m/2}\frac{\d^m P_n(x)}{\d x^m} \label{Pnm}\\
Q_n^m(x)&=|1-x^2|^{m/2}\frac{\d^m Q_n(x)}{\d x^m} \label{Qnm}
\end{align}
for all $x\neq\pm1$. Some authors multiply by these by $(-)^m$ for $|x|<1$. For negative $m$:
\begin{align}
	P_n^{-m}(x)&=(-)^m\frac{(n-m)!}{(n+m)!}P_n^{m}(x) \\
	Q_n^{-m}(x)&=(-)^m\frac{(n-m)!}{(n+m)!}Q_n^{m}(x) .
\end{align}

The Legendre functions of the second kind can be expressed as
\begin{align}
Q_n^m(x)=P_n^m(x)Q_0(x)-W_{n-1}^m(x), \label{QPQ0R}
\end{align}
where 
\begin{align}
W_{n-1}^0(x)=\sum_{k=1}^n\frac{P_{k-1}(x)P_{n-k}(x)}{k},
\end{align}
and $W_{n-1}^m$ may be defined by the same recurrence relations over $n$ and $m$ as do $P_n^m$ and $Q_n^m$ (with $n\rightarrow n+1$). For example they can be created for $m>0$ by  
\begin{align}
W_{n-1}^{m+1}=\frac{(n-m+1)W_n^m-(n+m+1)xW_{n-1}^m}{\sqrt{|1-x^2|}}.
\end{align}
Unlike $P_n^m(x)$, $Q_n^m(x)$ is non-zero for $n<m$, with $Q_n^m=-W_{n-1}^m$. 

Like for $m=0$, the $Q_n^m$ functions for $|x|>1$, i.e. the spheroidal harmonics, should be computed via a backward recurrence scheme if one needs $n\gtrsim 7$. 

\subsection{Negative degree}
\label{secQnmneg}

For negative degree $n$ with $-m<n<0$, $Q_n^m(x)$ is finite but cannot be defined from differentiating the $m=0$ functions as in Eq.~\eqref{Qnm} since $Q_n^0$ is not defined for $n<0$.
Instead, $Q_n^m(x)$ may be calculated from the terminating hypergeometric series in Eq. 14.3.12 of Ref.~\cite{NIST:DLMF}, explicitly for $0\le n<m$:
\begin{align}
Q_{-n-1}^m(x)=&\frac{(m+n)!(m-n-1)!(2n-1)!!}{(1-x^2)^{m/2}}\nonumber\\
&\times\sum_{q=0}^\frac{m+n}{2}\!\frac{(-)^{q+n}~x^{m+n-2q}}{(m+n-2q)!2q!!\prod_{k=0}^{q-1}(2n-2k-1)}. \label{Qn<0ex}
\end{align}

For any pair of indices $n,m$ (with interchange $n\rightarrow -n-1$ or $m\rightarrow-m$), there are two independent solutions to Legendre's equation. So $Q_{-n-1}^m(x)$ are linearly related to $Q_n^m$ and $P_n^{-m}$, (combining relations from Ref.~\cite{tables2014}):
\begin{align}
Q_{-n-1}^m(x)-Q_n^m(x) = (m+n)!(m-&n-1)!(-)^n P_n^{-m}(x),\nonumber\\ 
&-m\leq n<m.\label{Legendre m>n}
\end{align}
Perhaps of practical interest, $P_n^{-m}(x)$ is only infinite at $x=-1$ (the negative $z$-axis).

Explicit expressions for low degrees are (for $|x|<1$):
\begin{align}
Q_0^0(x)&=\text{atanh}(x) = \frac{1}{2}\log\frac{1+x}{1-x} \\
Q_{-1}^1(x)&=\frac{x}{\sqrt{1-x^2}} \\
Q_0^1(x)&=\frac{1}{\sqrt{1-x^2}}\\
Q_1^1(x)&= \sqrt{1-x^2}\text{atanh}(x) + \frac{x}{\sqrt{1-x^2}} \\
Q_{2}^1(x)&=3x\sqrt{1-x^2}\text{atanh}(x) + \frac{3x^2-2}{\sqrt{1-x^2}}\\
Q_{-2}^2(x)&=\frac{3x-x^3}{1-x^2 }\\
Q_{-1}^2(x)&=\frac{1+x^2}{1-x^2 }\\
Q_{0}^2(x)&=\frac{2x}{1-x^2 }\\
Q_{1}^2(x)&=\frac{2}{1-x^2 }\\
Q_{2}^2(x)&=3(1-x^2)\text{atanh}(x) - \frac{3x^3-5x}{1-x^2 }.
\end{align}

\section{Series coefficients relating logopoles and PSSHs}\label{sec beta}
In Sec. \ref{secLvsQP} it was deduced that for $m>0$ logopoles are expressed as an infinite series of PSSHs:
\begin{align}
L_n^m=&\sum_{p=m}^\infty \beta_{np}^m Q_p^m(\bar{\xi})P_p^m(\bar{\eta}).
\end{align}

For $m=1$, from inspection for small $p$, the coefficients appear to reduce to
\begin{align}
\beta_{np}^1=
\begin{cases}
-\frac{2(2p+1)}{p(p+1)}\big(1-\frac{n!(n+1)!}{(n+p+1)!(n-p)!}\big) & p\leq n \\
-\frac{2(2p+1)}{p(p+1)}. & p>n \\
\end{cases}
\end{align} 
For larger $m$ the expressions get more complex. The sum over $k$ in \eqref{LvsQPdouble} is numerically unstable, producing large errors for $p\gtrsim 20$. 

For $n=-m$, from inspection of small $p$, it seems that
\begin{align}
\beta_{-m,p}^m=\frac{2\left(1+(-)^{p+m}\right) (2 p+1)}{(m-1)!(p-m+1)(p+m)}.
\end{align}


By comparing the source integrals for logopoles (transformed to fit on the line segment O''O') and PSSHs, it is clear that $\beta_{np}^m$ are also the coefficients of the following expansion:
\begin{align}
2^{1-n}\frac{(v+1)^{n+m}}{(1-v^2)^{m/2}}= \sum_{p=m}^\infty \beta_{np}^mP_p^m(v),
\end{align} 
and by the orthogonality of the associated Legendre functions,
\begin{align}
\beta_{np}^m=\frac{(p-m)!}{(p+m)!}\frac{2p+1}{2^n}\int_{-1}^1 \frac{(1+v)^{n+m}P_p^m(v)}{(1-v^2)^{m/2}}\d v,
\end{align}
which may be somehow used to find a stable expression for $\beta_{np}^m$.

\section{Deriving Eq. (\ref{L_m stable}), the stable form of $L_{-m}^{m}$}\label{stableproof}
Starting from Eq. \eqref{L_m vs Q} and substituting the coordinate transformations $u''=\frac{\xi\eta+1}{\xi+\eta}$, $u'=\frac{\xi\eta-1}{\xi-\eta}$, we get
\begin{align}
&L_{-m}^m=\frac{2^m(2m-1)!! e^{im\phi} }{((1-\eta^2)(\xi^2-1))^{m/2}}\sum _{k=0}^{m-1} \frac{(-)^k}{2 k+1} \binom{m-1}{k} \nonumber\\ &\times\!\frac{[\eta ( \xi^2\!-\!1) +\xi(1\!-\!\eta^2 )]^{2 k+1} -[\eta ( \xi^2\!-\!1) -\xi(1\!-\!\eta^2 )]^{2 k+1}}{\left(\xi ^2-\eta ^2\right)^{2k+1}}. \nonumber
\end{align}
The terms for each $k$ can be expanded in powers of $\eta ( \xi^2 -1)$ and $\xi(1 -\eta^2 )$ as
\begin{align}
&[\eta ( \xi^2 -1) +\xi(1 -\eta^2 )]^{2 k+1} -[\eta ( \xi^2 -1) -\xi(1 -\eta^2 )]^{2 k+1} \nonumber\\	&=2\sum_{q=0}^{k}\binom{2k+1}{2q+1} (\eta ( \xi^2 -1))^{2k-2q} (\xi(1 -\eta^2 ))^{2q+1}.
\end{align}
Then rearranging the summation order gives
\begin{align}
L_{-m}^m=	&2^{m+1}(2m-1)!!\left(\frac{1-\eta^2}{\xi^2-1}\right)^{m/2} e^{im\phi}\nonumber\\
&\times\sum _{q=0}^{m-1} \left(1-\eta ^2\right)^{2q+1} \xi ^{2 q+1} \sum _{k=q}^{m-1} \frac{(-)^k}{2k+1} \nonumber\\
&\times\binom{m-1}{k} \binom{2 k+1}{2 q+1}\frac{ \left(\eta  \left(\xi ^2-1\right)\right)^{2 k-2 q}}{ \left(\xi ^2-\eta ^2\right)^{2 k+1}}.
\end{align}
This is still unstable, but the sum over $k$ may be recognized as a Gauss hypergeometric function $_2F_1$, and the connection formula $_2F_1(a,b;c;x)=(1-x)^{c-b-a} _2F_1(c-a,b,;c;1-x)$  where $x=[\eta(\xi^2-1)/(\xi^2-\eta^2)]^2$ and $1-x=(1-\eta^2)(\xi^4-\eta^2)/(\xi^2-\eta^2)^2$ can be used to rearrange the sum over $q$ and extract  $m-2q-1$ factors of $(1-\eta^2)$, giving Eq. \eqref{L_m stable}.

%

\begin{acknowledgments}
We are very grateful to an anonymous referee whose thorough and insightful feedback has much improved the quality of the final manuscript. This work was supported by the MacDiarmid Institute for Advanced Materials and Technology (New Zealand).
\end{acknowledgments}

%

\end{document}